# Biophotons: A Hard Problem

L. De Paolis [1], R. Francini [2], I. Davoli [4], F. De Matteis [2], A. Scordo [1], A. Clozza [1], M. Grandi [5], E. Pace [1+], C. Curceanu [1], P. Grigolini [3], M. Benfatto [1, †]

1. Laboratori Nazionali di Frascati - Istituto Nazionale di Fisica Nucleare - Via E. Fermi 40 – 00044 Frascati (Italy);
2. Dipartimento di Ingegneria Industriale - Università di "Tor Vergata" - Via del Politecnico - 00133 Roma (Italy);
3. Center for Nonlinear Science, University of North Texas, Denton, TX 76203-5017, USA;
4. Dipartimento di Fisica - Università di "Tor Vergata" - Via della Ricerca Scientifica - 00133 Roma (Italy);
5. Istituto La Torre, Via M. Ponzio 10, 10141 Torino, Italy;

† Correspondence: M.B. maurizio.benfatto@lnf.infn.it ; E.P. elisabetta.pace@lnf.infn.it ;

**Abstract:** About a hundred years ago the Russian biologist A. Gurwitsch, based on his experiments with onion plants by measuring their growth rate, made the hypothesis that plants emitted a weak electromagnetic field which somehow influenced cell growth. This interesting observation remained fundamentally ignored by the scientific community and only in the 1950s the electromagnetic emission from some plants was measured using a photomultiplier used in single counting mode. Later, in the 80s several groups in the world started some extensive work to understand the origin and role of this ultra-weak emission, hereby called biophotons, coming from living organisms. Biophotons are an endogenous very small production of photons in the visible energy range in and from cells and organism, and this emission is characteristic of alive organisms. Today there is no doubt that biophotons really exist, this emission has in fact been measured by many groups and on many different living organisms, from humans to bacteria. On the contrary, the origin of biophotons and whether organisms use them in some way to exchange information is not yet well known; no model proposed since now is really capable of reproducing and interpreting the great variety of experimental data coming from the many different living systems measured so far. In this brief review we present our experimental work on biophotons coming from germinating seeds, the main experimental results and some methods we are using to analyze the data in order to open the door for interpretative models of this phenomenon and clarifying its function in the regulation and communication between cells and living organisms. We also discuss some ideas on how to increase the signal-to-noise ratio of the measured signal to have new experimental possibilities that allow the measurement and the characterization of currently unmeasurable quantities.

## 1. Introduction

All living systems emit electromagnetic radiation, a small number of photons of the order of about 100 ph/sec per square centimeter surface area, at least in the visible energy range. The scientific community calls this emission by the name of biophotons [1,2]. This emission is completely different from the normal bioluminescence observed in some organisms because it is present in all living organisms, from plants to human beings, and it is several orders of magnitude weaker. Biophotons cannot come from the contribution of thermal radiation in the visible energy range because a simple calculation using the Plack distribution tells us that the intensity of this latter radiation is several orders of magnitude smaller than the biophotons contribution [3]. Moreover, this emission ends when the organism dies; this excludes the possibility that it is the product of either some radiative decay produced by traces of radioactive substances present in the organism or by the passage of cosmic rays. The main characteristics of biophotons are, besides the very small intensity, a practically flat emission within the energy range between 200 and 800 nm and the fact that any type of stress due, for example, to some chemical agents or excitation by light, induces a very fast increase in the emission up to a few order of magnitude, followed by a relatively slow decrease to the normal values with a non-exponential law [1,2].

In the 1920s the Russian biologist A. Gurwitsch [4], doing experiments with onion plants by measuring their growth rate, observed that this was strongly influenced by the fact that the various seedlings were close or not and that this behavior persisted, even if the possibility of any bio-chemical exchange had been eliminated. On this basis he hypothesized the existence of a sort of "morphogenetic field" responsible for the regulation of cellular growth in living systems and capable of influencing the mitotic activity of surrounding tissues. He called this weak emission "mitogenetic radiation". Despite the confirmation of his results (see the paper by Gabor and Reiter [5]), the scientific community forgot completely the Gurwitsch's results and then his work faded in the background. Only thirty years later, with the development of photons detector technology, Colli and Facchini [6,7] made the first measurement of electromagnetic emission coming from living organisms. This work was taken up again in the 1980s by F.A. Popp [2] and co-workers who started extensive work to understand more in detail the origin and the meaning of biophotons emission.

Despite the wealth of experimental results, the questions of what biophotons are, how they are generated, and how they are connected to life are still open. There are two hypotheses [1,2]. The first sees the emission as the random radiative decay of some molecules excited by metabolic events, like, for example, oxidative process and radical reactions, in the cells. The second scenario assigns the emission to a coherent electromagnetic field generated within and between the cells by some biochemical reactions in which, perhaps, oxygen atoms are involved. Both theories predict that any type of perturbation generated by non-specific stress gives rise to an increase in emission as experimentally observed. The two hypotheses are not mutually exclusive and the experimentally revealed emission could have dual origin.

At the same time, there is experimental evidence that such radiation carries important biological information [8-11]; for example, the radiation emitted by growing plants or organisms can increase by as much as 30% the cell division rate in similar organisms, the so-called mitogenetic effect [11-13].

Recently, biophotons are also revealing themselves as a non-invasive method for research in biology from applications in toxicology [14] to human health monitoring [15] as well as identification and treatment of diseases, especially cancer [16].

In this short review we deal with bio-photon emission arising during the germination process of seeds of various kinds. The experimental setup we refer to measure the biophoton emission is based on photomultiplier techniques, and it is essentially constituted by a dark chamber and a photomultiplier sensitive to the visible energy range. The detector works as a photon counter, and the experimental data are the number of photons detected within a well-defined time window. In this way, the experimental data are essentially a time series where the counts detected in the chosen time window are reported as a function of the time calculated from the moment of closure of the experimental setup [1,6,7,17]. The duration of the experiment can vary from a few hours to many days, depending on the germination time of the considered seeds.

We have recently published a paper [17] in which the time series generated by the biophotons emitted during the germination of lentil seeds has been analyzed using the diffusion entropy analysis (DEA) method. This method [18,19] is based on the concept of complexity developed by Kolmogorov [20] and is evaluated through a scaling index $\eta$, which is expected to depart from the ordinary value $\eta = 0.5$ if the signal shows some degree of anomalous complexity. The complexity of the time series is derived through the evaluation of the Shannon entropy associated with the diffusional trajectory [17–19] obtained from the experimental time series. The main result of Ref. [17] is that the biophoton emission shows conditions of anomalous diffusion with a substantial deviation of the scaling coefficient from the ordinary value $\eta = 0.5$ throughout the duration of the experiment. At the beginning of germination, the condition of anomalous diffusion is due to the presence of so-called crucial events, i.e., situations in which the system's memory is reset to zero. As the seeds germinate and roots and leaves begin to develop, the type of complexity associated with the experimental data completely changes its nature, and the departure from the condition of random diffusion is due to the so-called fractional Brownian motion (FBM) [21] regime.

Biophotons can be regarded as an index of thermodynamic activity [2] and changes in emission rates over time (usually hours) results in changes in scaling parameters [17]. When biophoton emission rates are used in conjunction with an analytical technique like DEA, they have the potential to document dynamic changes of complexity in a developing organism or complex adaptive system. From this point of view the germination process of lentil seeds could be seen as a process that has phase transitions accompanied by changes in patterns of complexity (crucial events to FBM condition). At the beginning of germination there is a process of cellular differentiation, which leads to the development of leaves and roots, and this phase could require the presence of crucial events detected in the biophotonic emission. In other words, it cannot be excluded that the type of criticality [11] inherent in the germination process requires a form of phase transition not yet known. These results offer the possibility to investigate the existence of variation in complexity patterns in a variety of different developing organisms and provide evidence for the importance of the exchange of information (entropy) transfer for cell-to-cell communication during organismal development [11].

It is interesting to note that Mancuso and collaborators [22,23] use the concept of swarm intelligence with reference to the network of roots generated by plants living in natural conditions. The presence of crucial events in the initial stage of germination could have something to do with the birth of this amazing radical intelligence.

In this paper, we present a brief review of our work on spontaneous emission coming from seeds during the germination process with some ideas for developing the experimental set-up to increase the signal-to-noise ratio with the aim of having access to information that is difficult to obtain today. A paragraph is also dedicated to a brief description of the main applications of biophotons in the life sciences.

## 2. Methods and Experimental Data

Our experimental setup was formed by a germination chamber, a photon counting system and a turning filters wheel [17]. A drawing of the experimental setup we are using to measure the spontaneous emission of germinating seeds is shown in Fig. 1.

The photon counting device is a H12386-210 high-speed counting head (Hamamatsu Photonic Italia S.r.l, Arese (MI), Italy) powered at +5 Vcc. The phototube is sensible in the wavelength range between 230 and 700 nm with a peak sensitivity at 400 nm [24]. The data acquisition and control of the experiment is done via an ARDUINO board and a computer equipped with a LAB-VIEW system (National Instrument, Austin, TX, USA).

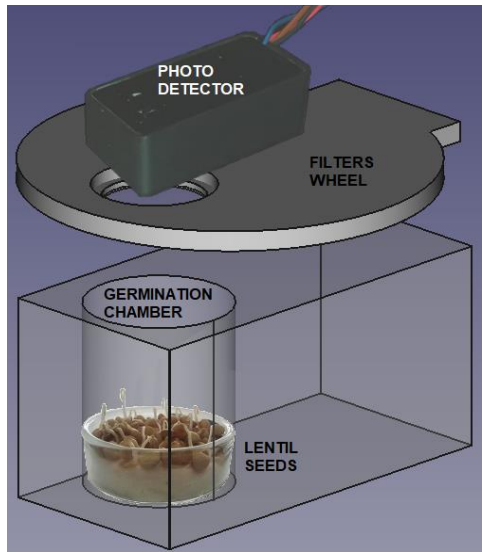

**Figure 1.** Schematic view of the experimental setup used in our experiment. The germination chamber is built with black PVC to avoid any contamination of light from outside.

The whole experimental set-up works as a single counting system and the detector can see a single photon with just the quantum efficiency of the photomultiplier. The acquisition time window is fixed at 1 s and within this window the entire system has a dark count of approximately 2 counts/sec, perfectly in line with the data sheets of this specific photomultiplier which indicates 1.7 counts/sec [25]. A turning wheel holding a few long pass glass color filters is placed between the germinating seeds and the detector. The wheel has eight positions. Six are used for the color filters, one is empty and the last one is closed with a black cap, see [17] for details.

Seeds are kept in a humid cotton bed placed in a Petri dish; they are normal seeds bought in a supermarket. Without any seed the emission consists in a monotonic decreasing tail due to the residual luminescence of the material, a consequence of the light exposure of the experimental chamber. The emission tail arrives in few hours at the dark counts value.

A typical measured signal with seeds and the wheel in the empty position is displayed in Fig.2. This signal (green points) refers to the emission of 76 lentils for a total duration of 72 hours from the moment of closure of the experimental apparatus after insertion of the Petri dish with the seeds into the measurement chamber.

The initial behavior (few hours) is dominated by the residual luminescence. The germination-triggered biophoton emission clearly emerges about 5-7 hours after closing the chamber (the change in the slope of the counting curve), and then it becomes dominant with a signal well above the detector dark counts (red points) for the entire duration of the experiment.

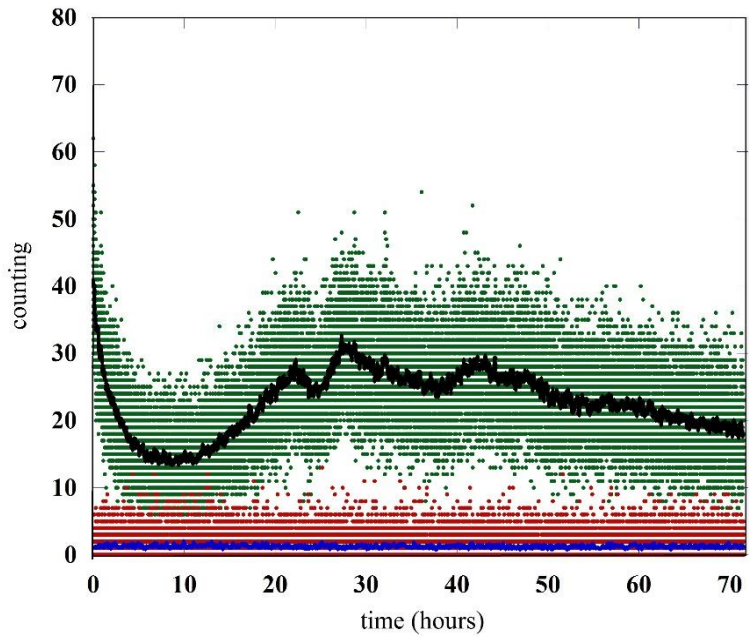

**Figure 2.** Comparison between the signal generated by the germinating seeds and the signal in the dark condition. The raw data are the green (seeds) and red (dark count) points. The black and blue curves are the raw data (count/sec) averaged over one minute.

The shape of the temporal evolution of the biophotons emission reported in Fig. 2 looks to be quite a general feature in the germinating phase of seeds. For example, the emissions of common wheat (Triticum aestivum) [26] and of seeds of Arabidopsis thaliana [27] are very similar to those presented here. This is quite interesting and needs a deeper discussion.

The comparison [28] between the emission of the 76 lentils and that of a single bean is reported in Fig. 3. In both cases, the emission was activated by the watering process and analyzed in a wide time interval ranging from the end of the residual luminescence until the time when germination generated roots and leaves. The time scales of the 76 lentils are completely different from the time scale of the single bean. For this reason, to highlight the common characteristics of the two emissions, we rescaled the time scale of the single bean by a factor of 0.164. In this way, it was possible to align the emission maxima of the two cases, the C peaks in the figure. The two curves have been moved further to have the zero of the time scale positioned in the first minimum. This means that the values 10 and 100, respectively, for the lentils and the single bean have been subtracted from the original time scale. To have the same number of counts in peak C, the values of the counts relating to the single bean were multiplied by a factor of 2.28.

The time rescaling procedure used here is based on the popular logistic equation [28, 29] which can be used in different systems to describe the growth of a population which reaches the final steady-state value which is specific for any system. The logistic equation played an important role in biology, contributing at the same time to the emerging science of chaos.

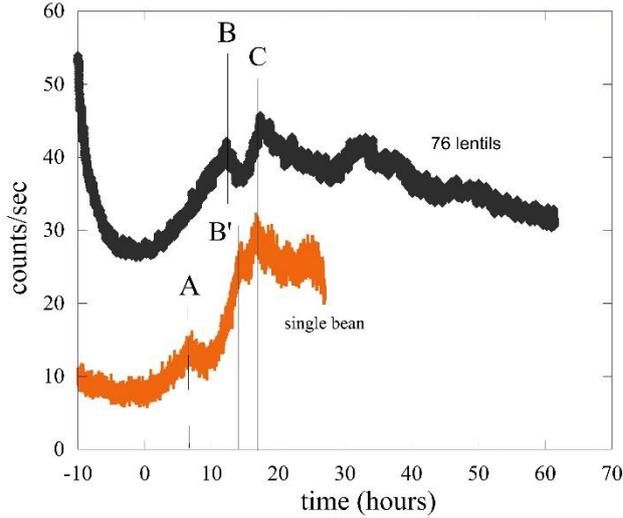

**Figure 3.** Comparison between the biophoton emission of the single bean with the emission of the 76 lentils. The two curves are counts per second averaged over 1 minutes. For clarity, the curve relating to the emission of lentils has been moved upwards, and it has been used in the time scale of the lentil's emission (see text for details). The capital letters in the figure indicate the main emission peaks observed in the experimental data.

Here the logistic equation takes the form:

$$\dot{n}(t) = a \cdot n(t) - b \cdot n^2(t) \tag{1}$$

where n(t) can be thought as the number of cells growing because of watering the seeds [30] and the numbers a and b are constants that depend on the system. The solution of Eq. (1) can be written as:

$$n(t) = \frac{a \cdot C e^{at}}{1 + b \cdot C e^{at}} \tag{2}$$

and C depends on the initial conditions n(0) through the relation: $C = \frac{n(0)}{a - n(0) \cdot b}$. We make the conjecture that the rate of biophoton emission is proportional to the derivative of the number of cells, i.e. to $\dot{n}(t)$:

$$\dot{n}(t) = \frac{a^2 \cdot C e^{at}}{(1 + b \cdot C e^{at})^2} \tag{3}$$

Cells can be thought as a kind of interacting units in the living organism, for a single type of unit the time derivative will reach a maximum at a time determined by the parameters a, b and the initial conditions. The corresponding emission has a regular trend with only one maximum reached at the time $t_{max} = \frac{1}{a} \ln\left(\frac{1}{\beta}\right)$ and intensity $I_{max} = \frac{a}{4b}$. The number β is defined as $\beta = b \cdot C$.

We hypothesized that the saturation time of the ordinary logistic equation in different systems corresponds to the maximum emission peak to a certain extent, and we rescaled the time scale of the bean so that its maximum photon emission rate coincided with that of the lentils. In this way we can make a comparison between them reducing the time of the slower growth by a reducing factor, 0.164 in the case of the bean.

The fact that the emissions of different seeds had a very similar temporal behaviour led us to hypothesize the existence of a sort of generalized logistic equation as a universal property of the connection between the system growth and the photon emission.

The experimental data show a wealth of structures with a succession of maximums and minimums distributed throughout the duration of the experiment. In details, between time 0 and 20 h, the lentil emission presented two peaks (B and C) separated by about 5 h; the same two peaks (B' and C) were present in the emission of the single bean, but here they were separated by about 14 h. The biophoton emission of the bean showed a further peak (peak A) at about 43 h (these values in the bean time scale) after the minimum position at zero time in the scale. This peak was absent in the lentil emission. It should also be noted that in the germination phase, between zero and peak C, the growth phase of the emission presented at least two slopes.

This is a clear evidence of the presence of different type of units in the seeds that could be activated at different time with different time scales. The simplest generalization of equation (3) that considers the presence of different type of units and with which to make a phenomenological fit of biophotonic emission as a function of time could be of the type:

$$\dot{n}(t) = a^2 e^{at} \sum_{i=1}^{J} \frac{C_i}{(1 + b_i \cdot C_i e^{at})^2} \qquad (4)$$

where the different constants could be determined by a fit procedure done by using the experimental data. In Fig. 4 we present such a type of analysis by comparing the experimental data relating to the emission of the 76 lentils with two fits made using equation (4) with J=1 and J=5.

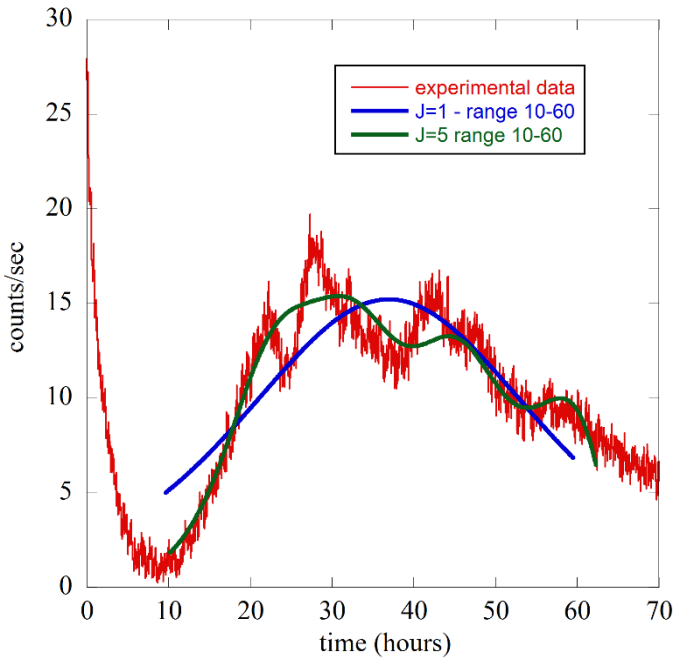

**Figure 4.** Comparison between the biophoton emission of the 76 lentils (red line) with two fits using Eq. 4 of the text with J=1 (blue line) and J=5 (green line). The experimental data are counts per second averaged over 1 minutes. The time scale is now the original one used in the raw experimental data presented in Fig.1

The two fits are done using the experimental data in the time range 10-60 hour. We do not report the values obtained from the fit procedure, both for brevity and because this analysis essentially aims to give a qualitative indication on the path to followed for the development of models capable of shedding light on the mechanisms underlying the generation of biophotons.

It seems quite clear that only fits made based on many-component functions can reproduce, at least qualitatively, the shape of the time series representing the measured experimental data, supporting in this way the idea that the germination process can also be thought of as an activation of different cell groups at different time characteristic for each group. From this point of view, each unit can be thought of as nodes in a network where each node is able to interact with its neighbors and make choices based on the increase or decrease in global benefit [31]. The system spontaneously evolves towards criticality, leading at the same time to the emergence of cooperation and intelligence. By intelligence we mean here the fact that a local interaction has changed into a long-term one making the single units sensitive not only to their nearest neighbors but also to the units very far away from them. Analysis of single bean experimental data produces the same type of fits and leads to the same type of interpretation, which is why they were not shown in this work.

For a full understanding of the biophoton phenomenon one also needs to consider the processes of excitation of molecules which will then give an emission during the radiative decay process. All the details of cellular environment, i.e., spectral distribution of the density of states, are important and can influence excitation as well as emission of the molecules, and the processes related to biophoton emissions should be essentially considered as connected to the non-thermal states occurring in the living cell. The processes in non-living substances are mostly related to thermal equilibrium states at room temperature, and thus they cannot lead to the molecular excitation and consequently to emission of the biophotons in optical region.

There are many models in the literature in this regard (see the reference [32] for a review), essentially based on three processes. The first is DNA replication (it is considered as a possible source of biophotons) during the cell cycle, when the activation energy of DNA replication may be used to excite molecules, prior to cell division that includes the membrane formation. The second is the synthesis of ATP that occurs during cellular metabolism, when the ion fluxes through membrane (particularly in the mitochondrion) and the last one based on the oxidation process of some biological molecules in the living system.

**2. Statistical analysis of the experimental data**

In this paragraph we briefly present the main results regarding the statistical properties of the time series representing the experimental data using the probability distribution function approach and the Diffusion Entropy Analysis methods.

*2.1. The Probability Distribution Function approach*

The first part of this paragraph consists of a brief description of the method to determine the probability distribution function $P_m(T)$ from experimental data. This can be done by counting how many times we measure a number m of photons for a given acquisition time window T. In a semiclassical picture of the optical detection process the phototube converts the continuous cycle-averaged classical intensity $\bar{I}(t)$ in a series of discrete photocounts. Thus, the number m of photocount obtained in an integration time T is proportional to the intensity of the light that arrives on the detector [33]. A photocount experiment consists of a sufficiently large number of measurements of the number of photocounts in the same integration period T. The result of the measurement is expressed by the function $P_m(T)$ which represents the probability of obtaining m counts in the acquisition time window T. This function obtained from the experimental data is analysed by determining mean, variance and other moments of higher order with the aim of highlighting the properties of the experimental $P_m(T)$. The purpose is to determine some statistical properties of emitted light through the properties of the distribution function, considering that, at least in some particular cases, there is a direct correspondence between the statistical properties of the light, some characteristics of the physical process that is producing the measured light and functional form of the $P_m(T)$. We do not repeat here the entire theoretical derivation that leads to the determination of $P_m(T)$, but we want to remember that there are only some cases that lead to an analytical form of this function. Details can be found in the references [28,33].

The simplest case is a stable classic light wave where the cycle-averaged intensity has a fixed value independent on the time [33]. In this case the distribution has a Poissonian form like:

$$P_m(T) = \frac{\langle m \rangle^m}{m!} e^{(-\langle m \rangle)} \qquad (5)$$

where $\langle m \rangle = \xi \bar{I}$. For Poisson distribution the variance is equal to the average $\sigma^2 = \langle m \rangle$. Any departure from the Poisson distribution could be an indication of a non-classical nature of the light and can be measured by the Fano factor [34] $F$ defined as $\sigma^2 = F \langle m \rangle$.

A Poisson distribution is a sign of a system in coherent states, in this case quantum states correspond to classical electromagnetic waves [28,33,34], but, at the same time, this distribution also occurs for experiments where the integration time $T$ is much longer than the characteristic time of the intensity fluctuations of the light beam.

The photocounts distribution can be also derived for a complete thermal chaotic light [28,33,34]. In this case the distribution takes the form:

$$P_m(T, M) = \frac{(m + M - 1)!}{m! (M - 1)!} \left(1 + \frac{M}{\langle m \rangle}\right)^{-m} \left(1 + \frac{\langle m \rangle}{M}\right)^{-M} \qquad (6)$$

where ⟨m⟩ is the average number of photons and M is the number of field modes. Thermal states are classical and there is the following relation between average and variance:

$$\sigma^2 = \langle m \rangle + \frac{\langle m \rangle^2}{M} \qquad (7)$$

In general, the coefficient M can be huge, this means that the variance becomes almost equal to the average value, and we find the same relationship valid for the Poisson distribution. As consequence, for large M the thermal photocount distribution approaches the Poisson distribution. This implies that it is difficult to discriminate between coherent and thermal states when many modes are present, in agreement with the discussion of Ref. [34] using this approach.

In reference [17,28] we have already presented a detailed analysis of the experimental data relating to lentils and single bean. For brevity, we present here only the comparison between the $P_m(T)$ obtained from the experimental data of lentil seeds related to the time window between 20 and 30 hours (original time scale of Fig. 1) and two fits done using equations (5) and (6). The results are presented in Fig. 5 The two fits have essentially the same $\chi^2$ and there is no reason to prefer one over the other.

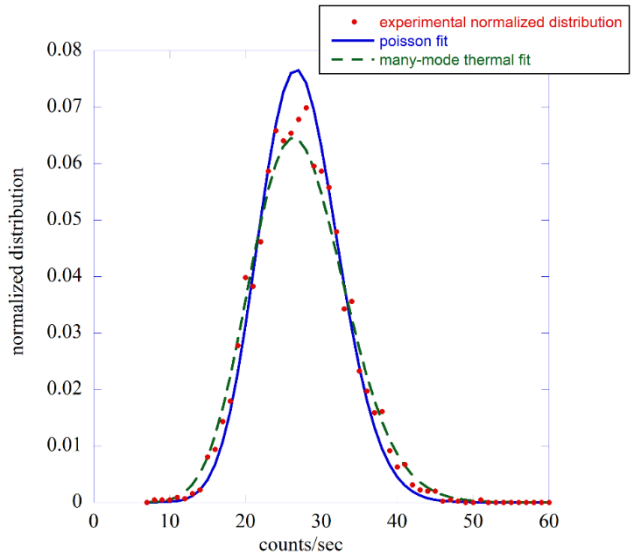

**Figure 5.** Comparison between the experimental count probability distribution function (red points) relative to lentils emission with two fits using a Poisson function (solid blue line) and a many-mode thermal function (dashed green line). The emission period is between 20 and 30 hours.

In this case, the experimental average count is $\langle m \rangle = 27.25$ and the variance is $\sigma^2 = 34.01$. The Fano factor $F \sim 1.25$ indicates a photocount statistics of super-Poissonian type. The Poissonian type of fit gives a value of the average counts equal to $\langle m \rangle = 27.18 \pm 0.07$, while the multi-mode thermal function gives $\langle m \rangle = 27.27 \pm 0.05$ and $M = 65.0 \pm 0.03$, in this last case, Equation (7) is roughly satisfied. This result confirms what we have found in previous works [17,28]. The experimental photocount distribution function always has, at least for the time windows considered in our experiment, a variance greater than the average value, indicating a super-Poissonian type of behaviour, that is typical of either thermal emission or emission with a very short coherence time compared to the time window of the measurement. This makes very difficult to discriminate between coherent and thermal states using this type of analysis, in agreement with the discussion of Ref. [34]. Clearly, the possibility of proving whether the biophotonic emission of living beings is coherent and measuring some parameters such as the coherence length and time is extremely important for the implications that this type of findings would have in defining the role that coherent processes have in biology. Coherence parameters can be measured by light interference or light correlation functions [34,35]. The non-classical nature of the emitted light could be assessed by the measure of the higher order correlation functions associated with the electromagnetic field [33,35]. This type of measurements is extremely challenging with the experimental set-up used up to now because we face at the same time with signal of very low intensity and coming from non-stationary processes.

*2.2. The Diffusion Entropy Method*

There is now clear evidence that biological systems cannot be described by the ordinary prescriptions of equilibrium statistical mechanics. This means the need to have analysis methods that can highlight all the deviations from the canonical form of equilibrium to understand the breakdown of the conditions on which Boltzmann's view is based: no memory, short-range interaction and no cooperation. Any deviation from the canonical form is a measure of the system complexity. There is still no unanimous consensus on the origin of complexity. In our case, the idea that complexity emerges from self-organization seems the most appropriate. The seed can in fact be thought of as a system that self-organizes when it begins to germinate because of watering. In general, a complex system is formed by several interacting units generating a whole with specific properties such as non-linearity, self-similarity, self-organization, just to quote a few. The complexity can be thought as a delicate balance between order and randomness, and when one of the two prevails complexity turns into simplicity.

In the complexity literature there exists a wide consensus on the importance of Kolmogorov complexity [20] and especially of the Kolmogorov-Sinai entropy [36]. The evaluation of the Kolmogorov complexity has been subject of extended literature. A dedicated discussion of this literature is out of the scope of this paper, we refer to the discussion presented in Ref. [19] for details. These authors illustrated two research directions aiming at evaluating Kolmogorov complexity, one called compression aiming to establish the Lyapunov coefficient directly and the second one called diffusion, which is based in converting the original time series into a diffusion process. The Kolmogorov complexity is turned into a scaling factor η, that is expected to depart from the ordinary value η = 0.5.

The technique of analysis used here is based on the diffusion approach and it is called diffusion entropy analysis., the so-called DEA analysis. This method was introduced into literature in the early 2000s [37,38] and it is based on converting the experimental time series, like the emission we record with our experimental set-up, into a diffusional trajectory. The complexity of the signal is determined through the evaluation of the Shannon entropy associated to the diffusional trajectory under the assumption that the complexity of the signal may be revealed by the anomalous scaling of the diffusional trajectory.

The time axis is divided into bins of size s (in our case s=1 sec) and we assign to the n-th bin the value ξ(n), which is the number of photons emitted in that small-time interval. For notation simplicity, the time series is considered as a continuous-time signal ξ(t), in this way the diffusional trajectory can be defined as

$$x(t) = \int_0^t \xi(t')dt' + x(0) \tag{8}$$

It is convenient to consider the $x^2(t)$ time series because it is directly related to the correlation function of the original time series [17,37,38]. The scaling properties are determined through the long-time limit behavior of the correlation function $\langle \xi(t_1)\xi(t_2) \rangle$ and the average can be made over a large number of realizations of $x^2(t)$ using the moving window method. See references [37,38] for details. Following the standard approach of assuming that the correlation functions are stationary, it is possible to define a normalized correlation function totally independent of the absolute values of $t_1$ and $t_2$:

$$\Phi_\xi(\tau) = \frac{\langle \xi(t_1)\xi(t_2) \rangle}{\langle \xi^2 \rangle} \tag{9}$$

where $\tau = |t_1 - t_2|$ and it is related to the $x^2(t)$ time series by the equation:

$$\langle x^2(t) \rangle = 2 \langle \xi^2 \rangle \int_0^t dt' \int_0^{t'} dt'' \, \Phi_\xi(t'') \tag{10}$$

We can now relate the complexity of ξ(t) to the anomalous scaling of the diffusion trajectory x(t). Using the Fractional Brownian Motion and Hurst notation [21,39] we indicate the scaling factor with the symbol H. Differentiating Eq. (10) twice with respect to the time and supposing that $x \propto t^H$ we get:

$$\Phi_\xi(t) \propto 2H(2H-1) \, t^{2H-2} \tag{11}$$

which, when H deviates from the ordinary value H=0.5, has, in the long-time limit [40], the structure $\Phi_\xi(t) \propto \pm \frac{1}{t^\delta}$ with δ=2-2H. Any deviation from the value H=0.5 indicates an anomalous behaviour of the time series which therefore presents some type of complexity even in the case of a stationary regime.

Although the conversion of the time series $\xi(t)$ into a diffusion trajectory leads naturally to relate the complexity of $\xi(t)$ to the Hurst coefficient $H \neq 0.5$, there exists another source of anomalous behaviour [41] of the diffusion trajectories that cannot be described through stationary correlation functions.

One of the key features of most complex systems is the presence of the so-called renewal events [41,42]. The happening of a renewal event resets the memory of the system and the sequence $\tau_i$ of the waiting times between successive renewal events are completely uncorrelated and independent. Renewal events generate a rejuvenation of the system, giving rise to a dynamic where whenever an event of this type occurs the system restarts from a completely new state. Renewal events characterized by the fact that the time interval between successive events is described by a waiting-time probability density function which has the important asymptotic properties $\psi(\tau) \propto \frac{1}{\tau^\mu}$, with µ ranging from 1 to ∞. Crucial events are renewal events corresponding to the condition $1 < \mu < 3$.

A classic example of a crucial event is the sudden change of direction in the flight of a swarm of birds. When such an event occurs, the global velocity of the swarm vanishes and the birds fly in a new direction which has no correlation with the previous one [43,44]. These types of events are not confined only to the swarms of birds but can be found in many biological and physiological processes [45] and play an essential role in the self-organization process of the living system and in keeping it healthy [46].

The scaling evaluation implies that we observe many equivalent but distinct realizations of the same diffusional process. For the analysis of data generated by the emission of bio-photons we have at our disposal only one time series. We generate the diffusional trajectory $x(t)$ according to the earlier prescription and, in order to make the statistical analysis, we convert this diffusional trajectory into many realizations such as to make it possible to do an ensemble average. These realizations are performed through a window of size $l$ that we move along the trajectory $x(t)$. Assuming a window of length $l$ ranging from $t$ to $t + l$, the value $x(t)$ is the initial position of the random walker that goes in a time $l$ from the origin to a value $x = x(t + l) - x(t)$. The different trajectories of the same time length $l$, needed to calculate the probability distribution function $p(x, t)$ and the related Shannon entropy, can be generated changing the initial position of the random walker, i.e. the t value.

This method allows us to find the anomalous scaling associated to the experimental data, but it is not able to discriminate whether the scaling factor is due to stationary or non-stationary correlation functions. For this purpose, the DEA algorithm must be modified by introducing the concept of stripes. Rather than converting the original experimental data into a diffusion process $x(t)$ directly, we divide the ordinate axis into many bins of size s and record the times at which the experimental signal crosses the border of two neighbouring stripes. In this way we obtain a new time series $\{t_i\}$. At any of these times an event occurs. We replace the experimental time data with a time series z(t) defined as follows. If time $t$ coincides with one of the times $t_i$, we set $z(t) = 1$ and $z(t) = 0$ otherwise. In other words, if the time $t$ corresponds to the occurrence of an event, the random walker makes a step ahead by the fixed quantity 1. The diffusion trajectory can be now obtained using again Eq. 8 with the surrogate time series $z(t)$. This is a very short description of the DEA approach with and without stripes; details can be found in Ref. [38,47].

To distinguish the anomalous scaling generated by crucial events from other types of anomalous scaling, we denote it with the symbol η rather than H. It is possible to demonstrate [41] that there are several different equations relating η and µ depending upon its value. In detail: $\eta = \mu - 1$ for $1 < \mu < 2$; $\eta = \frac{1}{\mu-1}$ for $2 < \mu < 3$ and $\eta = 0.5$ for $\mu > 3$. If complexity is generated by crucial events, the region $\mu > 3$ corresponds in any cases to stationary fluctuations and is interpreted as the manifestation of ordinary equilibrium statistical physics, conversely the region with $\mu < 3$ is the land of non-stationary behaviour, either temporary when $2 < \mu < 3$, or permanent for $1 < \mu < 2$.

The experimental data show extreme variability in terms of observed intensity, for this reason we decided to divide the total acquisition time of about 72 hours of into six regions, the first five having a length of 10 h while the last one is greater, being equal to 22 h. The idea behind this way of analysing data is to understand if µ changes with time during the germination process, what is its value at different stages of germination and how these compare with the values obtained in the analysis of some other physiological processes, like in heartbeats [45] and brain dynamics studied by EEG recording [48].

In figure 6 we show the six regions chosen for the analysis, separated by vertical black lines. In each of these regions we used DEA analysis with and without stripes in order to determine the various scaling factors. A similar analysis was also done for the dark counts (red dots in Fig.2) in order to have a comparison with experimental data coming from a case without any seeds.

We have found that dark count yields the ordinary scaling, thereby showing that no temporal complexity of either kind may occur in the absence of any seed in the chamber. In the presence of seeds in the chamber anomalous scaling emerges. The analysis with no stripes yields a scaling significantly larger than the scaling obtained by DEA supplemented with stripes. Furthermore, while the scaling factor remains practically constant in all the six different regions when obtained by using the DEA without stripes, there is a significant time dependence for the scaling factor obtained by the DEA with stripes. In the first phase of germination, within the three first temporal regions, the value of μ is significantly less than 3, a value which increases with the progress of the germination process. Without stripes the $\eta$ value goes from 0.694 (μ=2.44) to 0.796 (μ=2.25), while with stripes it goes from 0.496 (μ=3.01) to 0.596 (μ=2.67). For brevity we do not report here the complete analysis (see Ref. [17] for details), but we present a short summary using an average procedure on the different scaling factors. These numbers related to the different temporal regions are divided in two sets, in the first one the average is performed using the scaling factors obtained in temporal regions #1-3, while in the second set the average is made using the scaling factors coming from the regions #4-6 (see Table 2 in Ref. [17] for the numbers).

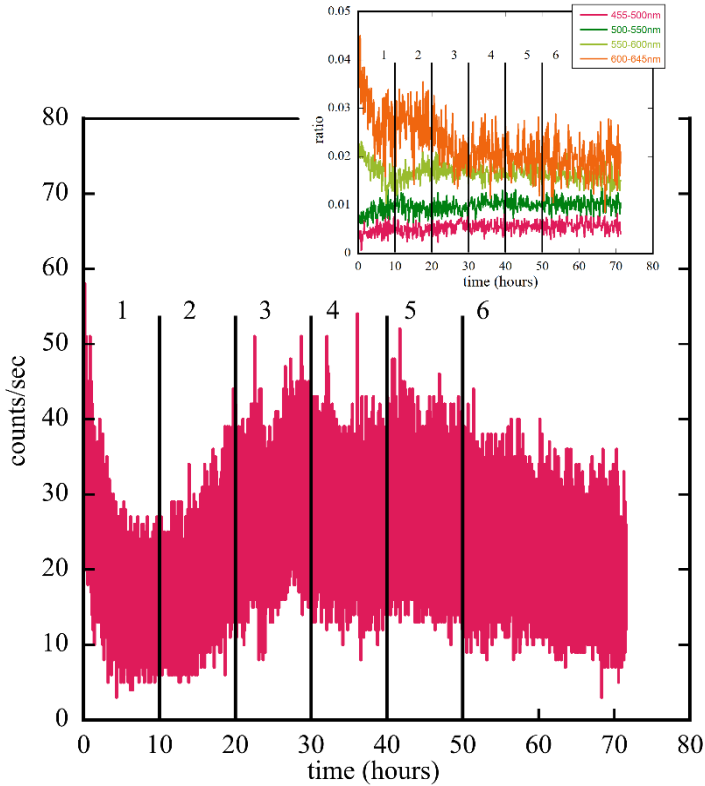

**Figure 6.** Number of photons emitted during the germination of lentil seeds (violet points). The black lines represent the six different regions used for the DEA analysis. The different spectral components of the signal relating to four spectral windows are also indicated with different colors in the inset at the top of the figure.

Using DEA without stripes the average scaling index related to the first 30 hours (regions #1-3) is η=0.77±0.03 which corresponds to a mean μ=2.30±0.05, while in the second three regions (regions #4-6), the average scaling index goes to η=0.72±0.02, corresponding to a mean μ=2.39±0.04. Essentially there is not a time dependence of the scaling factor throughout the whole germination process. In contrast, when applying the DEA with stripes, the results showed a clear and significant time dependence. In fact, in the first three temporal regions the same analysis produces a mean scaling factor η=0.56±0.04 which corresponds to a mean μ=2.79±0.11, while the last three regions give a mean scaling factor η=0.50±0.01 which corresponds to a mean μ=2.99±0.03.

These results clearly indicate that in the first three temporal region the departure from the condition of random diffusion is due to the presence of crucial events while the FBM regime dominates in the last stage of seed growth. In other words, it seems that during the germination process the non-ergodic component tends to vanish with time and complexity becomes dominated by the stationary infinite memory. However, the DEA analysis without stripes gives scaling factors significantly larger than that achieved by the use the DEA with no stripes. This indicated the fact that the germination process generates both crucial events and non-crucial events of FBM type. Therefore, the complexity of the emission of biophotons can be thought of as a mixture between two dynamics, one associated to crucial events and the second to non-crucial events of FBM type having infinite memory, the latter can be interpreted as a form of quantum coherence, which becomes predominant in the late germination phase.

In the inset of Fig. 6 (the right upper part of the figure) we show the different spectral components of the signal relating to the various wavelength intervals as shown in the figure. The experimental data were obtained through the use of long-pass glass color filters, and the curves corresponds to the average number of photons in each wavelength interval divided by the total signal without any filter. The complete analysis and a detailed description of the method can be found in the Ref. 28. It is clear that the different ratios change as a function of time, according to the moment of germination. In particular the high-energy components (green and violet curves) remain constant for the entire time of the measurement, while the lower-energy parts change in relative intensity; in details the orange part decreases at the beginning for few hours, it remains constant for up to hours 20 and then it slowly decreases until reaching a constant value after hours 30. This behavior is associated with a simultaneous increase in the yellow-green component of the spectrum in the hours 10-30. After this time interval all spectral components remain essentially constant until the end of the experiment. It is interesting to note that hour 30 represents the border between the time region where complexity is due to presence of crucial events and that instead dominates by FBM type of fluctuations. In our opinion, all these results represent the first empirical data indicating that the germination process of lentil seeds is a process that presents phase transitions [11] accompanied by changes in patterns of complexity (crucial to non-crucial events).

This type of behavior can also be found in the heartbeats of patients under the influence of autonomic neuropathy [49]. The increasing severity of this disease has the effect of making μ move from the healthy condition close to μ=2 to the border with ordinary statistical physics μ=3, that correspond to a pathological state. In human beings, the presence of crucial events is a necessary condition for different organs to "talk" to each other, for example the heart [45] and brain [48], and it is also the condition for maintaining good health [46,49].

This may not be true for plants as they do not have well-defined organs, only at the beginning of the germination process is there a clear differentiation process which may require the presence of crucial events. However, in order to grow, plants need light to trigger the synthesis of chlorophyll which is necessary to produce the appropriate nutrients; this does not happen in our experimental apparatus which is a totally light-free environment. Therefore, the change in patterns of complexity may also be due to the beginning of a pathological process that will lead to the death of the plant.

Discriminating between these two hypotheses is of extreme importance because it could open up the possibility of using the emission of biophotons as a tool to understand the state of health of a living organism through the determination of scaling indices and therefore the presence or absence of crucial events. At the same time, as the emission of biophotons is a universal characteristic of living organisms, this study can lead to confirmation that the presence of crucial events is a necessary condition for a health status of any type of living organisms.

## 3. Biophotons in living materials

In this short paragraph we want to give a brief overview of the use of biophotons in recent studies in living matter, biophotons in fact are emitted by all living organisms. Since the Gurwitch's first studies, many groups, all over the world, studied, and are currently studying, this phenomenon in completely different contests. Hundreds of articles have been written. See for an extensive review Ref. [32].

### 3.1. Seeds and plants

Many groups are interested in studying germinating seeds for several reasons, mainly for testing germinal goodness of the seeds and to verify how the plant growing is affected by pests, pollution agents, insecticides, fertilizers, extreme atmospheric conditions. Assessment of the quality of the seed is one of the most essential tasks for seed certification: the study of the emitted biophotons by sample of seeds is a rapid method to assess the various seed quality parameters, as Sarmah states in Ref. [50].

Extensive studies about consequences of pests are performed at the Hungarian University of Agriculture and Life Sciences [51,52] observing in-vivo plants. The same group studies the effect of pollution agents, like, for example, cadmium on barley [53]. When pests are present and the insecticides are used, it is important to understand their effect on the plants. For example, a group of the Universidade Tecnologica Federal do Paraná (Brazil) is studying the Triticum aestivum treated by Thiamethoxam [54].

It is also important how fertilizers affect plants: Salieres's French group is studying how hydrogenated water acts on Medicago sativa plant affecting growth and development [55]. The Japanese Kobayashi's group studies azuki seeds and their behavior under heat shock [56].

Other studies are conducted for understanding which parts of the plants emit biophotons: the Mackenzie's British group separately observed the biophotons emitted by roots and by the upper part of the a-mung sprouts [57].

*3.2. Food quality*

The observation of the biophoton emission is widely used in this field because is a fast and non-invasive technique. Recently the Iranian-Brazilian group headed by Nematollahi produced an extended list of studies on food quality and food production quality [58].

For example, biophoton emission coming from chicken eggs have been used regarding the possibility for quality verification [59], while the emission from wine is used to test the different winery practices in France [60] and in Hungary [61] to improve the quality of the production.

Nowadays, food is stored for several days before arriving in our houses. Biophotons can help in this case too. Several groups have used this emission to understand the degree of freshness of several type of food in many different conditions [50,52,62] and the security of storage of food, like the work done by the Chinese group led by Gong [63].

Pulsed electric field technology is an important emerging modality for both biomedicine and the food industry: e.g. in medicine (electrochemotherapy, tissue ablation, novel methods for drug and gene delivery and therapy), in the food industry (pasteurization, food compounds extraction) and in biotechnology. Biophotons are used in this case for sensing of protein oxidation generated by pulsed electric field [64].

*3.3. Humans and animals*

Many studies have been conducted on human beings and animals. The production of biophotons inside the human body has been demonstrated by the Zangari's group observing a signature left by biophotons, by a technique based on the principle that ionic Ag+ in solution precipitates as insoluble Ag granules when exposed to light [65]. Other groups are trying to understand why and where these biophotons are produced [66-68].

Some studies have observed biophotons emitted by cell culture or tissue slices. Tumor cells displayed increased photon emissions compared to non-malignant cells: it is possible to use biophoton emissions as a non-invasive, early-malignancy detection tool, both in vitro and in vivo [69]. Mice synaptosomes' and brain slices biophoton emission for studying differences between Alzheimer's disease and vascular dementia, discovering that communications and information processing of biophotonic signals in the brain are crucial for advanced cognitive functions [70]

Both in vivo and in vitro studies have been done for evaluating the oxidative stress of human skin that is important for the skin cancer prevention [71-73].

Many studies are conducted in the Traditional Chinese Medicine (TCM). Recently Wang and his colleagues stated that "the study of the super weak glowing phenomenon of biology and death of people's thoughts, anger and death process is expected to promote the combination of Chinese and western medicine [74]. Guo and colleagues propose to study changes of ultra-weak luminous intensity of acupuncture points and meridians before and after needling stimulation [75]. Biophoton emission was measured at four sites of hands of type 2 diabetes patients, before treatment and after 1 and 2 weeks of treatment with TCM: the biophoton emission intensity decreased gradually with the course of the treatment [76].

The biophotons emission has been also correlated to mental states. An increment of spontaneous human biophoton emission caused by anger emotional states [77,78]. Stress levels can be detected making specific analysis of biophotons emission (in conjunction with other physiological parameters: this kind of assessment is under study for Ukrainian military personnel after frontline service [79].

**4. Future experimental upgrade and perspective**

For any experimental apparatus which aims to perform biophotonic measurements, possible improvements involve two approaches: increasing the number of detectors to enhance counts or enhancing the collection capacity. The first approach can turn out to be a double-edged sword. Biophotons are an endogenous production of a very small flux of photons, of the order of 100 ph/sec, within the energy range between 200 and 800 nm. Even in the darkest environment, a low background counter could have some spare counts per second. Increasing the number of detectors means increasing the sources of background and the covered solid angle, but the last feature is increased by a very few percentage, and the results turn out to be a disadvantage. In our opinion, enhancing the biophoton measurement capability of an experimental setup is best achieved by improving its collection capacity. In this section, we introduce and assess various upgrades aimed at increase the capabilities to collect luminescence biophotons in a scientific apparatus. Such improvements could also be applied to different apparatuses that aim to make measurements of rare low-background luminescence processes.

A first improvement is the introduction of the Fresnel lens in the apparatus. The Fresnel lens is a "sectioned" convex lens with vertical parallel planes forming concentric rings. Refraction bends the light rays and makes them converge into a single focus. Placing a Fresnel lens in front of the emitting source (germinating plants for example) luminescence photons can be focused to the counter (distances can be evaluated with a simulation), and thus the geometrical efficiency of the apparatus can be improved up to two orders of magnitude, as shown in Fig. 7.

A second improvement will be the building of an integrating sphere made with white Teflon. Teflon reflects more than 99% of the incident light in the visible energy range, so the light originally generated by the sample is reflected many times by the walls and finally reaches the hole where the photomultiplier is housed. In this way it collects the emission coming from every part of the sample and it is as if we had increased by factors the solid angle of measurement. The integrating sphere will be located inside a PVC black chamber to avoid any light contamination. (see the right panel of Fig.7).

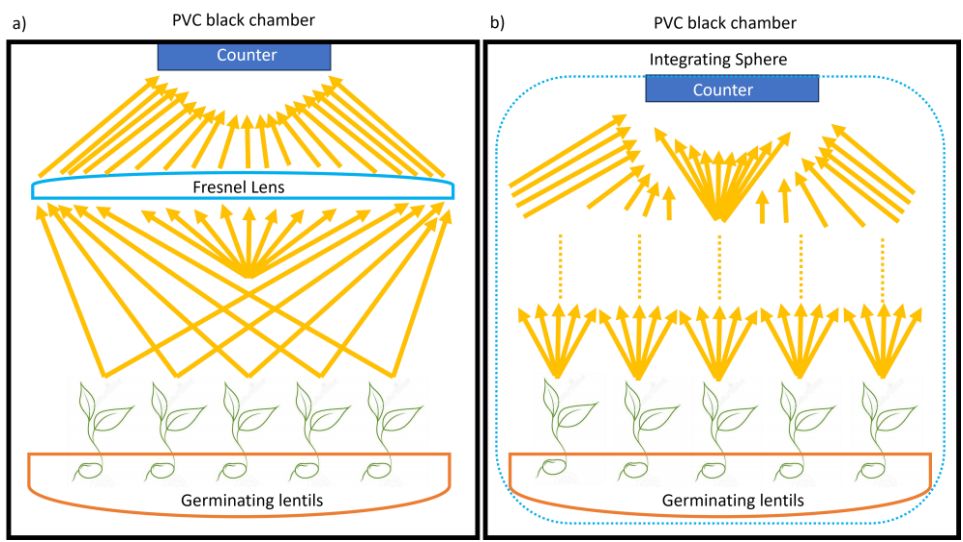

**Figure 7.** On the left: a) a schematic view of the apparatus with a Fresnel lens inserted to improve the number of light photons collected by the counter. On the right: b) a schematic view of the apparatus with the of the integrative sphere, made of white Teflon. The sphere allows the reflection of the light rays enhancing the collection capability of the apparatus.

The third improvement is the installation of light sources for calibration and validation testing of the experimental setup. Specifically, the best solution would be the installation of two light sources: one emitting coherent light and another emitting incoherent light. Light beams exhibit coherence when they combine as waves, involving the addition of amplitudes. On the other hand, they are considered incoherent when combined as particles, which entails adding the intensities, i.e., the square of the amplitudes, of the beams. Biophotons are Ultra-weak Photon Emission (UPE). Several well-crafted works provide an analysis of UPE concerning chaotic light fields. In contrast, numerous papers make assertions regarding coherent and squeezed states of UPE. The introduction of coherent and incoherent light sources in the apparatus is a fundamental step in investigating UPE properties through biophoton emissions.

So far, we mentioned the most upcoming improvements. However, several further improvements may be made. One is the installation of an infrared camera to get fundamental information about the growing cells. Such improvement can be tested also in lentils, but their chaotic vertical growth may produce an incomplete response collection. In cell cultures, where the layer is uniform, the infrared camera can collect more complete and clean information. Other improvements are the installation of devices and sensors to modify and monitor the temperature inside the chamber, the installation of irradiation systems, of variation of atmosphere or pH and constituents of the soil of the plants, to characterize the spectrometric response of the germinating seeds. In this way, the lentils experiment represents a pioneering study on biophotons. Apart from being a valuable source of information concerning plant growth, it also serves as a testing ground for evaluating devices and solutions applicable in the future, extending to the study of biophotons emitted by both healthy and cancer cell cultures [80,81,82].

## 5. Conclusions and suggestions for future

There is growing interest in the literature for the role that biophotons appear to have in biology, from studies on seeds and plants to those involving humans. In this short review we have essentially presented the results that our group has obtained in the study of seed germination processes, relating it to the results already known in the literature. We have seen a kind of phase transition during the germination process highlighted by changes in the complexity patterns (crucial and non-crucial events) and by a different behaviour of the spectral components. At the same time the analysis of the intensity and the shape of the emission in terms of a generalized logistic equation seems to indicate that during the germination period, the parts of the organism involved in the emission process change according to the degree of plant development.

In this contest the germination process can be thought as a spontaneous organization of a biological system generating non-equilibrium events and deviations from the dynamical processes of ordinary thermodynamics. The system evolution and the interaction between the different units leads to criticality [42], which therefore emerges directly from the processes of spontaneous organization of biological systems as a delicate balance between order and disorder.

The idea that biophotons, beyond the molecular mechanisms that generate them, are also a manifestation of the degree of complexity of the system can help us answer the question of how such a small signal can transmit information. This can be done using the complexity matching theory which was introduced to extend the Linear Response Theory of Kubo from the ordinary condition of thermodynamic equilibrium to the more general condition of perennial non equilibrium.

All living beings seem to emit biophotons and this emission is extremely sensitive to the "state" of the organism that is emitting them. Changes in biophoton emission spectra under any type of stress indicate that something happens in the living system, the changes in pattern complexity may be an indicator of a loss in cellular communication identified by crucial events [47] signalling a "loss of complexity," with the disappearance of crucial events, where loss of complexity may be an early sign of disease. From this point of view biophotons could have the role of establish links between the units of the complex living system, despite their ultra-weak intensity.

Following this line of thought we can hypothesize that biophotons are another communication route developed by nature to allow the exchange of information between different cells as well organisms. If this is true, we can imagine an active role for biophotons in the treatment of various pathologies, cancer or Alzheimer's, just to give a few examples, through the development of a system that emits biophotons of the right degree of complexity and intensity and which therefore can be "interpreted" by diseased cells to lead them through a healing process.

It is extremely important to increase the signal to noise ratio, and in this review, we have proposed an extremely economical possibility, but which could open the doors for experiments carried out on single seeds and which have a better energy resolution than the one we obtained. It should also be mentioned the possibility of using a CCD camera in experiments with plants and seeds to map the emitting parts as already done in the case of human being.

In this short review we have not presented anything about the so-called delayed luminescence experiments. These are done by studying the decay of the emission after the sample has been irradiated with an ultraviolet laser pulse with a typical width of 5ns and an energy of about 150mJ/pulse [83-85]. In this case, the return to normal emission occurs in a time that varies from a few tens of seconds to a few minutes, i.e. with a time scale much faster than one of the typical germination processes, which normally takes place over tens of hours. This indicates that there are essentially two types of emission, one associated with the relaxation of molecular species excited due to the normal metabolic processes of the living organism and the other originating from the relaxation of excited states induced by the external stimulus. Clearly, the two processes are closely connected, probably involving the same types of molecules, but, in our opinion, the decay channels and the processes behind the spontaneous emission are different.

A lot has been done and perhaps some answers are starting to emerge, but a lot still remains to be done, this requires work and imagination, but it is the fun part of the story.

**Acknowledgements** - We warmly thank the conversation we had with I. H. von Herbing, L. Tonello, and D. Lambert during the writing of this work.

**Funding**: We acknowledge support from the Foundational Questions Institute, FQxI, a donor advised fund of Silicon Valley Community Foundation (Grants No. FQXi-RFP-CPW-2008 and FQXi-MGB-2011) and from the John Templeton Foundation, Grant 62099. The opinions expressed in this publication are those of the authors and do not necessarily reflect the views of the John Templeton Foundation.